\documentstyle[12pt,aasms4,psfig]{article}
\def\infig#1#2#3{\epsfxsize=#3cm \centering{\mbox{\epsfbox{#2}}}}

\received{}
\revised {}
\accepted{}
\cpright{PD}{1996}

\lefthead{Vauclair \& Charbonnel}
\righthead{The Lithium Primordial Abundance}

\begin{document}
\title{ELEMENT SEGREGATION IN LOW-METALLICITY STARS
AND THE PRIMORDIAL LITHIUM ABUNDANCE}

\author{Sylvie Vauclair 
\footnote{Sylvie.Vauclair@obs-mip.fr}}
\affil{Laboratoire d'Astrophysique de Toulouse, CNRS UMR 5572, Toulouse,
France}
\and
\author{Corinne Charbonnel
\footnote{Corinne.Charbonnel@obs-mip.fr}}
\affil{Laboratoire d'Astrophysique de Toulouse, CNRS UMR 5572, Toulouse,
France}
\affil{Space Telescope Science Institute, 3700 San Martin Drive,
Baltimore,
MD 21218, USA}

\begin{abstract}

Observational constraints on the primordial lithium abundance are
important for the evaluation of the baryonic density of the universe.
Its precise determination however suffers from uncertainties due to 
the possible depletion of this element inside the stars.
Here we present and discuss new results on the lithium abundances 
in Pop II stars obtained
with most recent  stellar models including the best available physics.
We show that it is possible to account for the
general behavior of lithium observed in Pop II stars
without any free parameters. Macroscopic motions
inside the stars are needed but it is not necessary to specify their
exact nature to interpret the observational data.
This constatation allows to derive a parameter-free value for the
lithium primordial abundance :
$ \log  {\rm Li}_{p} = 2.35 \pm  .10$ in the $\log {\rm  H} = 12$ scale.

\end{abstract}

%\twocollumn

\section{INTRODUCTION}

Lithium-7 has been recognized for a long time as one of the four
primordial isotopes, which have been formed by nuclear reactions during
the first minutes of the Universe. Together with deuterium, helium-3 and
helium-4, its abundance is used as a constraint on the baryonic density
of the Universe. The fact that the observed abundances of these four
isotopes are consistent, within the uncertainties, with the same
baryonic density represents a success of the standard big-bang theory.
Recent improvements in the observational techniques (large ground-based
telescopes, Hubble Space Telescope, new image-processing techniques)
together with a better understanding of the chemical evolution of
galaxies should help reducing the uncertainties. At the present time
however, some ambiguities do remain which have to be solved for a better
understanding of the primordial universe (see Sarkar 1996 for a recent
review on the subject). 

In the present paper we address the question of the primordial $^{7}$Li
value. Its determination relies on the spectroscopic observations of
population II stars and on our knowledge of stellar physics. The lithium
abundances observed in the oldest stars with effective temperature 
larger than 5500K lie around $2$ x $10^{-10}$ in number compared to
hydrogen
(the ``Spite plateau", Spite \& Spite 1982). 
While cooler stars present a large dispersion with strong
evidences of lithium depletion, the abundance scatter in the ``plateau"
is very small except for a few stars with no lithium detection
(Thorburn 1994, Ryan et al. 1996, Spite et al. 1996, 
Bonifacio \& Molaro 1997).
This observational fact is often taken as an evidence that the 
average lithium value in these stars represents indeed the primordial 
abundance. 

Such a conclusion relies on a comparison of the observations 
with stellar models in which standard element segregation 
induced by gravitational and thermal settling 
is simply ignored while no other
process among those which could prevent this segregation is 
introduced.  
In stars, element segregation may indeed be 
counteracted by macroscopic motions like turbulent mixing or mass loss.
These motions can, in turn, lead to the nuclear destruction 
of lithium (Vauclair 1988, Pinsonneault et al. 1992). 

A realistic derivation of the primordial lithium must rely on a
consistent 
study of the internal structure of Pop II stars. 
In standard models, where nothing prevents the element segregation,
lithium is depleted at the surface of the stars due to the settling.
Its abundance increases with depth, goes though a maximum value 
(Li$_{\rm max}$) and then
decreases abruptly due to nuclear destruction.
Vauclair \& Charbonnel (1995, VC95)
discussed how this structure is modified in the presence of a
stellar wind. They showed that mass loss could rub out the lithium
depletion induced by the segregation and lead to a surface lithium
value close to Li$_{\rm max}$.
The average mass loss rate should lie around 10 to 30 times that of the
present solar wind. 
In this case the lithium behavior observed in Pop II stars
can be correctly reproduced.

Here we give more precise
computations of the internal structure of Pop II stars,
in which
we include better physics as in 
the recent solar models  tested by comparison with
helioseismology (Richard et al 1996, hereafter RVCD; Basu 1997).
We show that the maximum lithium value inside standard stellar models 
(Li$_{\rm max}$) is remarkably constant from star to star in the
effective temperature range of the Spite plateau. 
Varying the effective temperature, the metallicity or the mixing
length parameter never changes Li$_{\rm max}$ by more than 10 $\%$.
We give physical reasons for this behavior.

Our conclusion is that the constancy and robustness of this value may
represent
a clue for our understanding of the ``lithium plateau" in Pop II stars.
We suggest that the lithium abundance which is actually observed in
these stars is directly related to Li$_{\rm max}$.
Within this framework, a comparison of Li$_{\rm max}$ with the lithium
observations (Bonifacio \& Molaro 1997) allows to derive 
the primordial lithium abundance.

\section{SETTING THE STAGE}

\subsection{Observational constraints}

Lithium is observed in the atmospheres of stars with metallicities
ranging
from twice solar down to $10^{-4}$ solar. 
The upper envelope of the observed abundances follows a well defined 
trend :
in the $\log {\rm  H} = 12$ scale, 
the lithium upper values decrease from
about 3.3 for solar metallicity down to about 2.3 for metallicities 
ten times smaller and remain constant for still smaller metallicities
(Rebolo et al. 1988, Spite 1991).

When plotted as a function of the effective temperatures, for 
low-metallicity stars, the average lithium abundance does not vary 
significantly for effective temperatures larger than 5500K (the
``plateau").
For smaller effective temperatures, lithium is depleted due to nuclear
destruction.
According to various observers (e.g. Thorburn 1994; Molaro, Primas \&
Bonifacio 1995; Ryan et al. 1996), the plateau
is either horizontal or with a very small positive slope.
In any case, the small dispersion of the abundances
(Deliyannis, Pinsonneault \& Duncan 1993; Molaro et al. 1995; 
Spite et al. 1996; Bonifacio \& Molaro 1997) 
is a strong constraint on the type of processes which may have occurred
in these stars since their formation.
Furthermore the observations of the more fragile $^{6}$Li
isotope in a few halo stars (Smith et al. 1992, Hobbs \& Thorburn 1994),
if confirmed, represent an evidence that lithium has suffered
no, or only very small nuclear destruction in these stars.

\subsection{The physics}

In the radiative regions of stars, the various chemical species move 
with respect to the bulk of stellar gas,
because of the selective diffusion induced by the pressure, temperature,
concentration gradients, the electric field and the radiative
acceleration.
A self-gravitating gas-mixture cannot be in complete equilibrium and
remain
chemically homogeneous.
During the pre-main sequence phase, the stars are
convectively mixed, which forces homogeneity. When they become
radiative, the
elements begin to migrate from the initial stage towards a never-reached
equilibrium stage. This process creates an element segregation in stable
stars.

As a result, the abundances observed at the surface of the stars may be
different from the original ones. For main-sequence Pop I stars,
evidences of this effect have first been found in the so-called 
``peculiar A stars", where it can lead to variations up to several 
orders of magnitude (Vauclair \& Vauclair 1982).
More recently its signature has been found in the Sun from
helioseismology
(see RVCD and references therein).
Theoretical computations predicted abundance variations of the order of
10 to 20 $\%$ in the Sun due to this process. 
The fact that helioseismology confirmed it represents 
a great success for the theory of element segregation in stars. 

Macroscopic motions like shear-induced turbulence, rotational mixing or
mass loss may slow down the migration process and reduce 
the abundance variations. There are evidences both from
helioseismology and from the solar lithium abundance
that some mild motions must occur
below the solar convection zone (RVCD, Basu 1997). While these motions
reduce the slope of the
concentration gradients, they do not completely stop the segregation.

When comparing stellar models with the observations, all these physical
effects should be taken into account. However, while
pure segregation models are
parameter-free (except for the mixing length), 
macroscopic motions depend on the
stellar history and may be different for various stars
of the same mass and effective temperature. For this reason we
chose here to study
Pop II star models including element
segregation only. 
We show that such precise standard models give enough information 
to allow deriving a consistent
value of the lithium primordial abundance.

\section{ COMPUTATIONS AND RESULTS: THE LITHIUM ABUNDANCES}

The present stellar models were calculated with the Toulouse-Geneva
stellar evolutionary code (Charbonnel, Vauclair \& Zahn 1992)
in which improved microphysics was implemented since VC95.
Element segregation is treated as in RVCD, using
the diffusion coefficients obtained with the Paquette et al.
(1996) approximation.
We use the radiative opacities by Iglesias \& Rogers (1996),
completed with the atomic and molecular opacities by Alexander \&
Ferguson (1994).
The equation of state is described with a set of MHD tables 
(Mihalas, Hummer \& D\"appen 1988) specifically
calculated for the mass and metallicity domain we study here
(Charbonnel et al. 1998).
Stellar models of 0.80, 0.75, 0.70 and 0.65~M$_{\odot}$ 
have been computed from the pre-main sequence up to the turn-off,
with a metallicity [Fe/H] = -2 and a mixing length parameter $\alpha =$
1.6 (Figure 1). 
The effect on the Li profile of varying [Fe/H] and $\alpha$
has been tested for the 0.80~M$_{\odot}$ models (Figure 2).

The lithium profiles inside the stars are given in Figure 1 for three 
ages: 10, 12 and 14~Gyr.
The corresponding effective temperatures are given in Table 1, together
with the luminosities, the temperatures and densities at the base of 
the convection zone, and the surface and maximum lithium abundances.
For the models of  0.80, 0.75 and 0.70~M$_{\odot}$, lithium is depleted
in the convection zone due to the segregation. The abundance increases
inwards up to its maximum value, Li$_{\rm max}$.
Deeper in the star lithium is destroyed by nuclear reactions.
In the coolest 0.65~M$_{\odot}$ model, the two effects merge 
and no maximum appears.
In any case, none of these stars has kept its original abundance in
any part of the internal structure.

\begin{figure}
\infig{8cm}{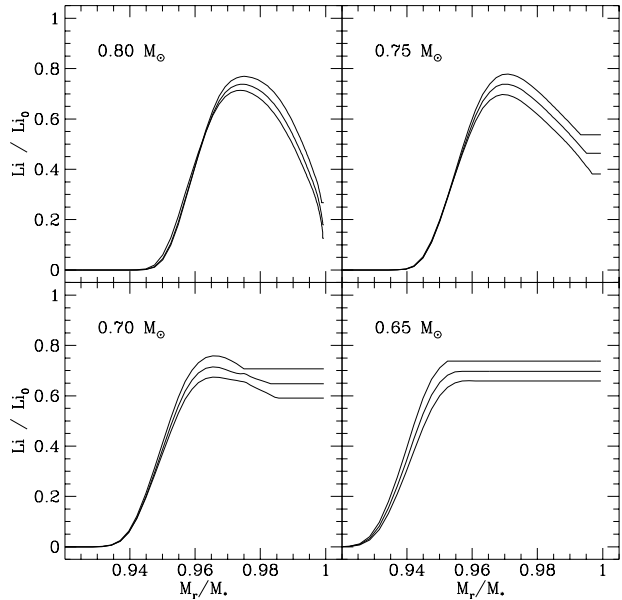}{8.8}
\caption{ Lithium profiles inside standard models for population II
stars, including segregation at three ages (10, 12, 14Gyr).
The ordinates are normalized to one and the abcissae are given in terms
of the mass fraction of the considered layer.
Lithium is depleted in the convection zones due to downward diffusion.
Its abundance increases with depth up to Li$_{\rm max}$ until it reaches
the nuclear destruction zone. 
While the surface depletion increases with increasing mass, the maximum
values inside the stars are nearly constant. \label{fig1} }
\end{figure}

\placefigure{fig1}
\placetable{table1}

The maximum lithium value inside the stars (Li$_{\rm max}$)
is remarkably constant from star to star. 
Figure 2 shows, for the case of a 0.8~M$_{\odot}$ star, the lithium
profiles obtained at 12 Gyr for three different values of the mixing
length 
parameter and of the metallicity (see also Table 2). 
For a given value of [Fe/H], the star has a deeper convection zone, and
consequently a higher surface lithium abundance when $\alpha$ increases.
On the other hand, decreasing [Fe/H] for a given value of $\alpha$ leads
to cooler effective temperature, deeper convection zone and less
important element segregation at the surface of the star. 
In spite of the changes in the surface lithium values and in the
effective 
temperatures of the model, the Li$_{\rm max}$ value never varies by
more than 10 $\%$.

\begin{figure}
\infig{8cm}{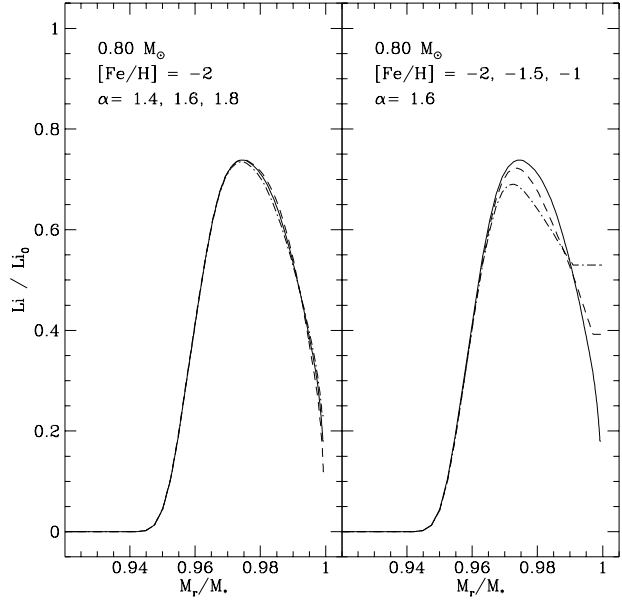}{8.8}
\caption{Left : lithium profiles at 12 Gyr inside a 0.8M$_{\odot}$ model 
computed with different values of the mixing length parameter $\alpha$
(solid line : $\alpha$=1.6, dashed line : $\alpha$=1.4, dashed-dotted line :
$\alpha$=1.8). 
Right : Same for different values of the metallicity (solid line :
[Fe/H]=-2, dashed line : [Fe/H]=-1.5, dashed-dotted line : [Fe/H]=-1). 
For a given value of [Fe/H], varying $\alpha$ only slightly changes the
surface abundance due to segregation. 
However, decreasing the value of [Fe/H] for a given value of $\alpha$
leads to a decrease of the effective temperature of the star at a
given age (at 12 Gyr, the 0.8M$_{\odot}$ model has 
effective temperatures respectively equal
to 6000, 6155, 6320K when [Fe/H] = -1, -1.5 and -2).
As a consequence, the base of the convection zone deepens, and element
segregation is less important at the surface of the star.
In any case Li$_{\rm max}$ never changes by more than 10 $\%$ \label{fig2}} 
\end{figure}

\placefigure{fig2}
\placetable{table2}

This remarkable result may be understood in the following way. In first
approximation, the lithium maximum appears 
at the place where the segregation and the
nuclear destruction time scales are of the same order. The segregation
time scale is defined as $\tau_S = $H$_p^2/$D where H$_p$ is the
pressure scale height and D the diffusion coefficient. The nuclear time
scale is defined as $\tau_N = (\lambda_{\rm Li,H} $N$_{\rm H})^{-1}$
where $\lambda_{\rm Li,H}$ represents the nuclear reaction rate per
particle
and N$_{\rm H}$ the number of protons per unit volume.
Above the lithium maximum $\tau_S << \tau_N $ and lithium is
depleted due to the element settling. Below the
maximum $\tau_S >> \tau_N $ and lithium is nuclearly destroyed. The
local values of $ \tau_N $ and $\tau_S $ only depend on the local
temperature and density. For this reason, the lithium maximum always
appears at about the same position inside the star, with 
about the same value.
It does not
depend on the depth of the convection zone. This explains the constancy
and robustness of the lithium maximum.

\section{DISCUSSION: THE PRIMORDIAL LITHIUM ABUNDANCE}

Stellar models including element segregation predict a surface lithium
depletion which depends on the effective temperature, the age and the 
metallicity of the stars.
This prediction is in 
contradiction with the observations of the lithium plateau
in Pop II stars.
However, as discussed above,
the lithium profiles in the standard models
present a maximum (Li$_{\rm max}$) which remains remarkably constant 
(within 10 $\%$)and
stable from star to star. 
This result leads to the idea that the observed lithium abundances may
be related to this maximum value.

When VC95 computed the effect of mass loss on the lithium abundances in
Pop II stars, they found the following characteristic behavior : for
very small mass loss rates, the surface lithium abundance is depleted
due to element settling. Increasing the mass loss rate contradicts the
settling and the surface lithium abundance first increases. It goes
through a maximum for a rate about ten times the solar wind and then
decreases for larger rates due to nuclear destruction. The 
most remarkable
result is that the maximum surface lithium abundance which is obtained
lies very close to Li$_{\rm max}$. 
This behavior has a simple physical reason related to the macroscopic
time scale (here
$\tau_{ML} = $H$_p/V_{ML}$ where $V_{ML}$ is the mass loss velocity
averaged over H$_p$) : the
maximum surface lithium value is obtained when the macroscopic 
time scale at the Li$_{\rm max}$ depth 
is of the order of the two other time scales
$\tau_{S}$ and $\tau_{N}$. 

The same behavior is expected for all kinds of
macroscopic motions : the surface
lithium value should always have a maximum of the order of Li$_{\rm max}$.
Since the observations of lithium in the plateau reveal a very
small dispersion around a stable value, this value must indeed lie
close to Li$_{\rm max}$. The shape of the lithium profiles as a
function of mass (Fig.1 and 2) shows indeed that small macroscopic
motions acting just below the convective envelope will on the average
give values close to Li$_{\rm max}$. 

In Figure 3, the observations compiled by Bonifacio \& Molaro (1997)
(BM97) are
shown for the plateau stars ($T{\rm _{eff}}  > 5500$~K).
In the same graph we have plotted the Li$_{\rm max}$ values obtained 
for 12 Gyr, $\alpha$ = 1.6 and [Fe/H] = -2. The
initial value is taken as A(Li) = 2.35 (where A(Li) represents the
logarithm of the lithium abundance in the log H =12 scale).
The comparison of the Li$_{\rm max}$ curve with the observed lithium
``plateau" allows to determine the primordial lithium value.

\begin{figure}
\infig{8cm}{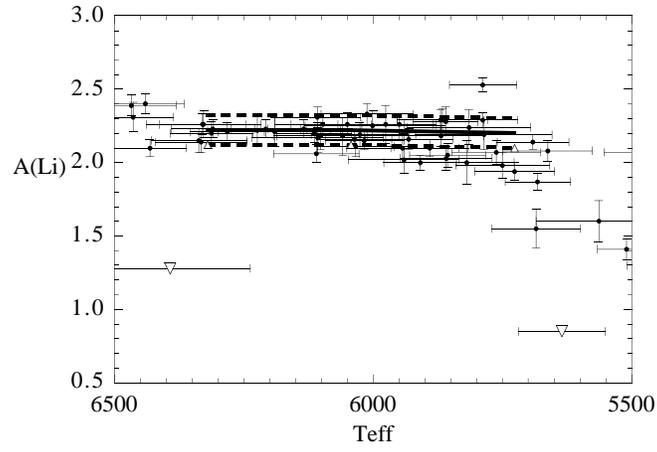}{8.8}
\caption{ Comparisons between the theoretical results obtained after
12 Gyr and the observations.
The observational points and error bars are from Bonifacio \& Molaro
(1997). 
The solid line shows the computed Li$_{\rm max}$ values; the dashed
lines represent the uncertainties of $\pm$.10 discussed in the text
\label{fig3}}
\end{figure}

\placefigure{fig3}
\placetable{table3}
\placetable{table4}

Table 3 gives the values of A(Li) as obtained without any correction 
from the analytical fits
of BM97 (their table 2). These values are precisely computed for the
effective temperatures given by our models for three masses and three
ages. 
As a further step, they are then increased by the logarithm of the corresponding
Li$_{\rm max}$/Li$_0$ value (given in Table 1). The new values are
labelled as
A(Li$_0$) in Table 4.

If the observed lithium was exactly Li$_{\rm max}$, all the A(Li$_0$)
values should be identical. In reality, we 
expect the 0.70~M$_{\odot}$ star to have suffer extra nuclear depletion,
as the bottom of the convection zone is very close to the nuclear
destruction layer. Also the age of these Pop II stars is supposedly  
larger than
10 Gyr. For these reasons, we chose to compare the four values given by
the 0.75 and 0.80~M$_{\odot}$ models at 12 and 14 Gyr.

Considering an uncertainty of about 10 $\%$ on the surface lithium
abundance compared to Li$_{\rm max}$, and taking into account the
systematic error in the observations quoted by Bonifacio \& Molaro (1997), 
we give for the primordial lithium abundance:
A(Li$_0) = 2.35 \pm 0.10$. 
When compared to BBN computations (e.g. Copi, Schramm \& Turner 1995) 
this result leads to a baryonic number between 1.2 and 5 $10^{-10}$.
For H~=~50, this value corresponds to 
$0.018 < \Omega _{b} < 0.075$.

\vspace{2cm}
We thank R.Cayrel for fruitfull comments on the manuscript. 
We are grateful to the Centre National Universitaire Sud de Calcul
for providing computer facilities for this study.
We thank the Institute for Nuclear Theory at the University of
Washington where this work was initiated.
C.C. acknowledges support provided by the Space Telescope Science
Institute.

\clearpage
\begin{deluxetable}{cccccccc}
\tablecaption{Parameters of the pure diffusion models ([Fe/H]=-2,
$\alpha$=1.6)\label{table1}}
\tablehead{
\colhead{M$_*$/M$_{\odot}$}
&\colhead{t}
&\colhead{T$_{\rm eff}$}
&\colhead{Log L/L$_{\odot}$}
&\colhead{T$_{bzc}(10^7$K)}
&\colhead{$\rho_{bzc}$}
&\colhead{Li/Li$_0$}
&\colhead{Li$_{\rm max}$/Li$_0$}
}
\startdata
0.65 & 10 & 5363 & -0.615 & 0.204 & 1.289 & 0.741 & \nl
     & 12 & 5392 & -0.589 & 0.197 & 1.110 & 0.700 & \nl
     & 14 & 5424 & -0.560 & 0.195 & 0.987 & 0.662 & \nl
0.70 & 10 & 5675 & -0.416 & 0.173 & 0.554 & 0.710 & 0.759 \nl
     & 12 & 5723 & -0.374 & 0.151 & 0.339 & 0.652 & 0.715 \nl
     & 14 & 5773 & -0.326 & 0.142 & 0.258 & 0.595 & 0.676 \nl
0.75 & 10 & 5978 & -0.205 & 0.117 & 0.108 & 0.547 & 0.778 \nl
     & 12 & 6044 & -0.137 & 0.105 & 0.066 & 0.465 & 0.738 \nl
     & 14 & 6087 & -0.050 & 0.087 & 0.032 & 0.384 & 0.680 \nl
0.80 & 10 & 6257 &  0.034 & 0.065 & 0.010 & 0.288 & 0.769 \nl
     & 12 & 6322 &  0.166 & 0.048 & 0.003 & 0.172 & 0.738 \nl
     & 14 & 6230 &  0.339 & 0.044 & 0.002 & 0.125 & 0.714 \nl
\enddata
\end{deluxetable}

\begin{deluxetable}{cccccccc}
\tablecaption{Parameters of the pure diffusion 0.8M$_{\odot}$ model at
12 Gyr for different [Fe/H] and $\alpha$ \label{table2}}
\tablehead{
\colhead{[Fe/H]}
&\colhead{$\alpha$}
&\colhead{T$_{\rm eff}$}
&\colhead{Log L/L$_{\odot}$}
&\colhead{T$_{bzc}(10^7$K)}
&\colhead{$\rho_{bzc}$}
&\colhead{Li/Li$_0$}
&\colhead{Li$_{\rm max}$/Li$_0$}
}
\startdata
-1      &1.6    &6000   &-0.011 &0.136  &0.078  &0.530  &0.691 \nl
-1.5    &1.6    &6157   &0.076  &0.091  &0.002  &0.392  &0.722 \nl
-2      &1.6    &6322   &0.166  &0.048  &0.003  &0.172  &0.738 \nl
-2      &1.4    &6243   &0.164  &0.044  &0.002  &0.119  &0.739 \nl
-2      &1.8    &6389   &0.166  &0.060  &0.006  &0.230  &0.735 \nl
\enddata
\end{deluxetable}

\clearpage
\begin{deluxetable}{cccc}
\tablecaption{A(Li) obtained from the analytical fits by BM97 for the
effective temperatures of our models on the ``plateau" \label{table3}}
\tablehead{
\colhead{A(Li)}
&\colhead{0.7 M/M$_{\odot}$}
&\colhead{0.75 M/M$_{\odot}$}
&\colhead{0.8 M/M$_{\odot}$}
}
\startdata
10 Gyr & 2.12 & 2.17 & 2.22 \nl
12 Gyr & 2.13 & 2.19 & 2.23 \nl
14 Gyr & 2.14 & 2.19 & 2.22 \nl
\enddata
\end{deluxetable}

\begin{deluxetable}{cccc}
\tablecaption{A(Li$_0$) derived when the correction Li${\rm max}$/Li$_0$
is applied to the analytical fits by BM97. See the text \label{table4}}
\tablehead{
\colhead{A(Li$_0$)}
&\colhead{0.7 M/M$_{\odot}$}
&\colhead{0.75 M/M$_{\odot}$}
&\colhead{0.8 M/M$_{\odot}$}
}
\startdata
10 Gyr & 2.24 & 2.28 & 2.33 \nl
12 Gyr & 2.28 & {\bf 2.32} & {\bf 2.36} \nl
14 Gyr & 2.31 & {\bf 2.36} & {\bf 2.37} \nl
\enddata
\end{deluxetable}

\clearpage \newpage

\clearpage \newpage

\end{document}